\documentclass[a4paper]{jpconf}
\usepackage{graphicx}
\usepackage{subcaption}
\usepackage{hyperref}
\begin{document}
\title{First performance measurements with the Analysis Grand Challenge}

\author{Oksana Shadura$^{1}$, Alexander Held$^{2}$}

\address{${}^{1}$ University of Nebraska–Lincoln, 1400 R St, Lincoln, NE, United States}
\address{${}^{2}$ University of Wisconsin–Madison, 447 Lorch St., Madison, WI, United States}

\ead{oksana.shadura@cern.ch}

\begin{abstract}
The IRIS-HEP Analysis Grand Challenge (AGC) is designed to be a realistic environment for investigating how analysis methods scale to the demands of the HL-LHC. The analysis task is based on publicly available Open Data and allows for comparing the usability and performance of different approaches and implementations. It includes all relevant workflow aspects from data delivery to statistical inference.

The reference implementation for the AGC analysis task is heavily based on tools from the HEP Python ecosystem. It makes use of novel pieces of cyberinfrastructure and modern analysis facilities in order to address the data processing challenges of the HL-LHC.

This contribution compares multiple different analysis implementations and studies their performance. Differences between the implementations include the use of multiple data delivery mechanisms and caching setups for the analysis facilities under investigation.
\end{abstract}

\section{Introduction}

The Analysis Grand Challenge (AGC) started out as an integration exercise for IRIS-HEP~\cite{irishep}, connecting various areas of work within the institute and the surrounding ecosystem. It quickly transformed from that into a project that we positioned to be useful to the broader community. The AGC provides a testbed for physics analysis software developers to explore user experience, interfaces, and performance~\cite{Held:2022sfw}.

The goal of the analysis exercise is to demonstrate the handling of data pipeline requirements of the HL-LHC, including large data volumes, bookkeeping, and handling of different types of systematic uncertainties. It also includes investigations of the use of reduced data formats (e.g. PHYSLITE~\cite{physlite} or NanoAOD~\cite{nanoaod}), aligned with the goals of the LHC experiments. In addition to that, the project aims to engage users to explore columnar analysis concepts.

The AGC also aims to explore the concept of fast “interactive analysis” with a turnaround time of minutes or less. We are testing the feasibility of this idea by employing highly parallel execution in short bursts, which furthermore needs to happen with low latency. This also requires heavy use of caching to improve performance in subsequent executions of similar analysis tasks.

Another important target we envision for the AGC is to prepare analysis facilities for execution of analyses towards the HL-LHC and to provide new concepts and services for end users. 

\section{IRIS-HEP Analysis Grand Challenge components}
The AGC project includes work on a number of related items. While IRIS-HEP is involved in all of them, the project is structured in a way to allow for contributions on specific aspects of the project without having to interact with everything at once. Aspects of the work include:

\begin{enumerate}
\item defining a physics analysis task of realistic HL-LHC scope and scale, allowing to easily implement and re-implement it;
\item developing analysis pipelines that implement said physics analysis task;
\item finding and addressing performance bottlenecks and usability concerns for the pipelines implemented.
\end{enumerate}

These proceedings describe first performance measurements obtained, following the definition of an analysis task and its implementation with a specific pipeline as described in reference~\cite{Held:2022sfw}. We will briefly summarize the task and implementation here.

\subsection{AGC analysis task description}
The physics analysis task consists of a cross-section measurement of top quark pair production in final states with a single charged lepton. This task is chosen to capture relevant workflow aspects of a typical physics analysis. The analysis phase space is also somewhat generic, allowing to convert the setup into other types of analyses, such as searches for beyond the Standard Model physics phenomena. The analysis task features prominently the handling of different types of systematic uncertainties, including the handling of associated metadata and bookkeeping aspects. The analysis logic itself includes simple kinematic top quark candidate reconstruction.

The input data to this task is derived from Run-2 CMS Open Data~\cite{cms-open-data}, with around 400 TB available in MiniAOD format. The implementation described here makes use of inputs in an ntuple format, pre-converted from the MiniAOD format and consisting of about 1 Billion events (around 3.5 TB) made available publicly in XRootD-accessible storage at the University of Nebraska–Lincoln.

Open Data plays a crucial role in this project, as it allows anyone to participate without requiring specific access permissions. The analysis task focuses on demonstrating realistic workflows, but is not concerned about getting all physics details fully correct: it includes the use of made-up tools for calibrations and systematic uncertainties to probe the workflow aspects.

\subsection{An AGC implementation: software stack and analysis pipeline}

\begin{figure}[h]
\begin{centering}
\includegraphics[width=0.7\textwidth]{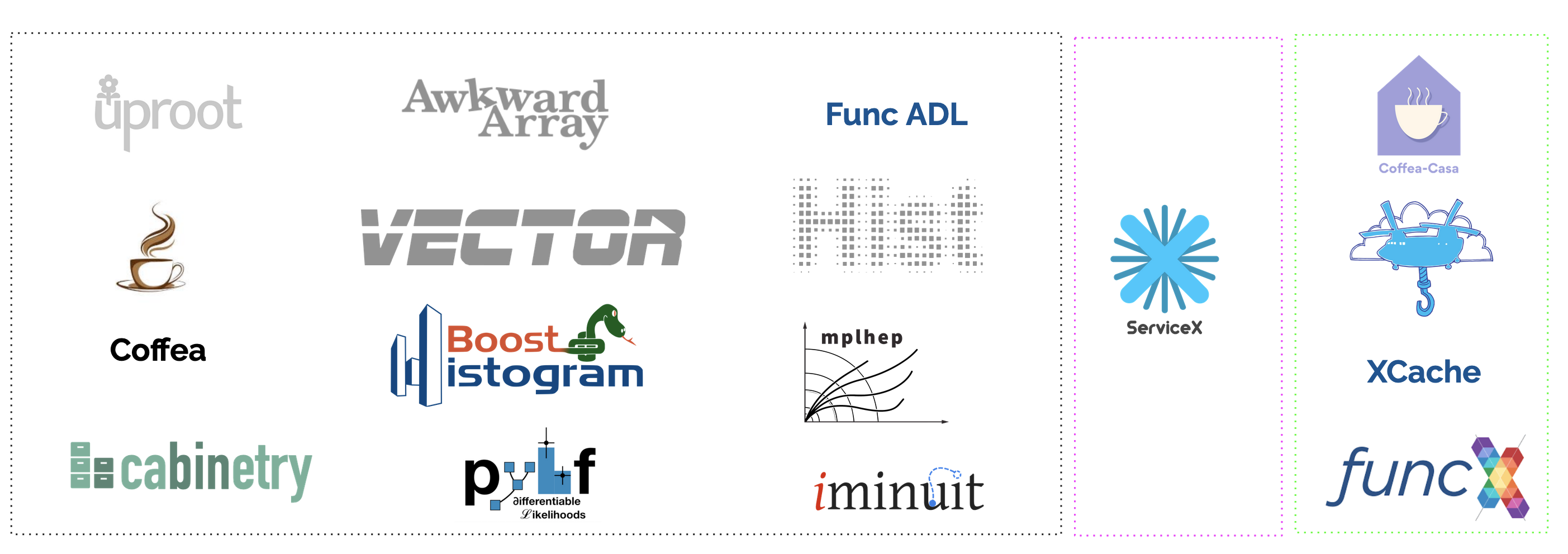}
\caption{Logos of software packages employed and considered in the AGC implementation described here, including packages focused on end-user physics analysis (left box), data delivery service (\textit{ServiceX}, middle box) and additional services provided by analysis facilities (right box).}
\label{fig:as}
\end{centering}
\end{figure}

The implementation of the AGC analysis task employed for the results shown in these proceedings makes use of a set of tools developed by IRIS-HEP and partners, with figure~\ref{fig:as} depicting the software stack being used and tested. 

The pipeline setup includes the \textit{ServiceX}~\cite{servicex} service providing the delivery of columns following a declarative \textit{FuncADL}~\cite{funcadl} request. The \textit{coffea}~\cite{coffea} framework then orchestrates distributed event processing and histogram production using \textit{uproot}~\cite{uproot} and \textit{awkward-array}~\cite{awkward}, with \textit{hist}~\cite{boost-histogram, hist} providing histogramming functionality. Visualization is provided by the \textit{mplhep}~\cite{mplhep} package. The statistical model construction is done by \textit{cabinetry}~\cite{cabinetry}, while statistical inference is performed using the \textit{pyhf}~\cite{pyhf,pyhf_joss} package.

Implementations for AGC analyses task are openly developed in the IRIS-HEP AGC repository~\cite{agc_code}, including the specific implementation used for the performance results in these proceedings. The repository also includes the categorization of datasets in terms of their role in the AGC analysis task and where to find them.

\subsection{An AGC implementation: R\&D on data management tools}

An ongoing research and development effort focuses on improved techniques for delivering physics events to analysts. This includes work on the development of dedicated data delivery services (such as \textit{ServiceX}) and integrating them together as one coherent ecosystem. All of this is intended to be available on analysis facilities, offering convenient user access and a good user experience.

We expect that the following projects can improve the performance of AGC implementations beyond the results shown in section~\ref{sec:results}:
\begin{enumerate}
\item XCache~\cite{xcache} — XRootD file-based caching proxy used for regional and site caches to store and serve datasets, helping to reduce latency and WAN traffic;
\item ServiceX — data extraction and data delivery service, offering “column-on-demand” functionality;
\item Skyhook DM~\cite{skyhook} — an extension of the Ceph distributed storage for scalable storage of tables and for offloading common data management operations (selection, projection, aggregation, and indexing, as well as user-defined functions).
\end{enumerate}

\subsection{An AGC implementation: R\&D on analysis facilities}

The HL-LHC will introduce new computing challenges surrounding the adaption of existing analysis paradigms at facilities to handle more data-intense end-user data analysis. To address this, there are ongoing efforts from different groups to study and prototype new facilities capable to assist in the upcoming physics analysis challenges.

\begin{figure}[!ht]
\begin{centering}
\includegraphics[width=0.7\textwidth]{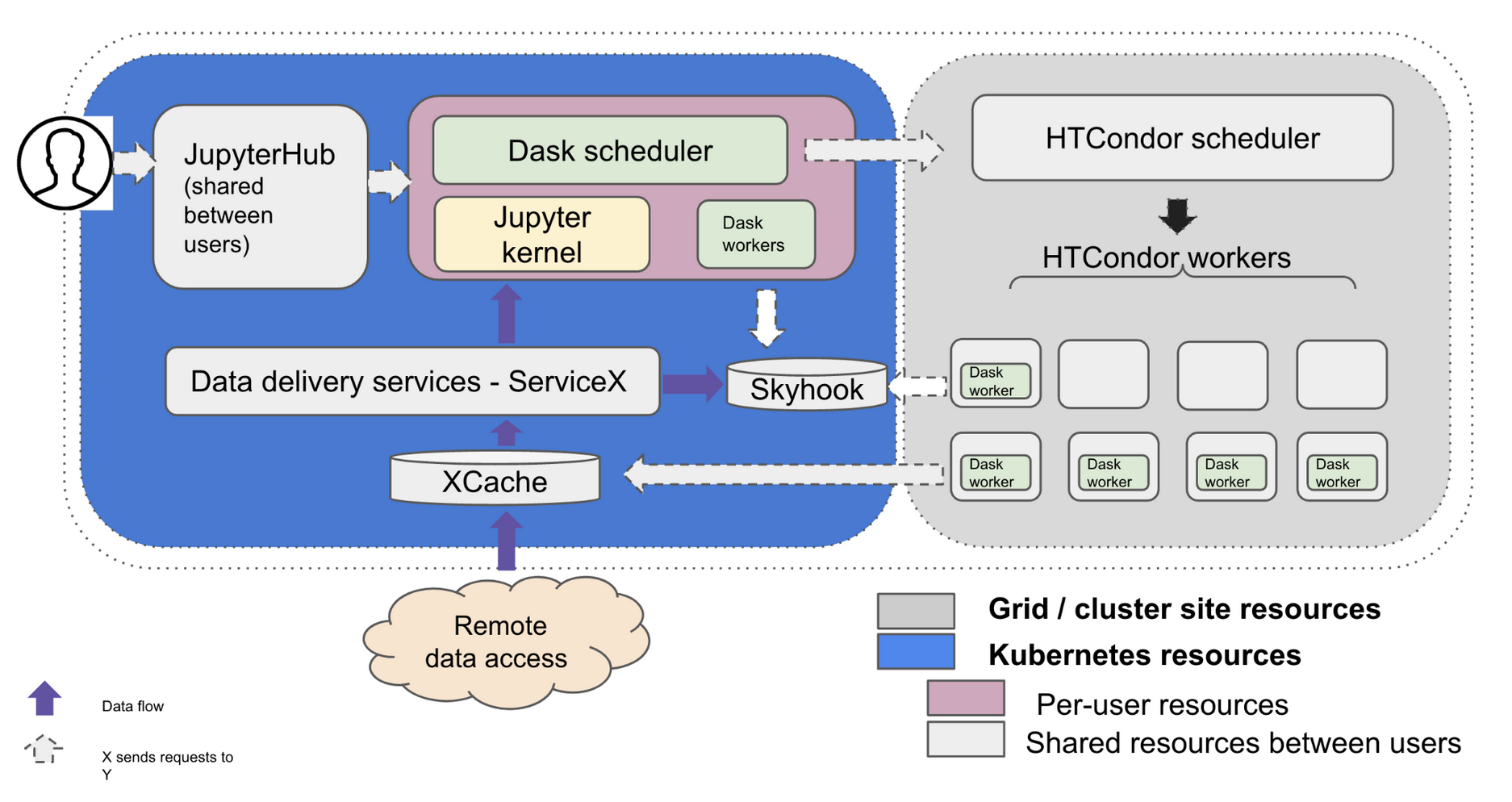}
\caption{Coffea-casa facility, developed by the University of Nebraska-Lincoln (IRIS-HEP).}
\label{fig:cc}
\end{centering}
\end{figure}

One of such prototype facilities, the Coffea-casa facility~\cite{cc}, developed by the University of Nebraska–Lincoln (IRIS-HEP), brings new, interactive paradigms for users from R\&D into production. This facility is used as a testbed for the IRIS-HEP Analysis Grand Challenge, offering the possibility for end-users to execute analysis at HL-LHC-scale data rates. This facility is adopting an approach that allows it to transform existing facilities (e.g. LHC Tier-2 sites) into composable systems, using Kubernetes as the enabling technology as described in figure~\ref{fig:cc}.

The Coffea-casa facility provides modularity and portability offering various configurations. The Coffea-casa team demonstrated the ability to port and customize the analysis facility setup to another site, co-locating it with the existing ATLAS Tier-3 analysis facility at the University of Chicago. The configuration required adjusting the Coffea-casa facility setup to become more Kubernetes-native, providing an HTCondor batch queue directly in Kubernetes.

\section{Results}
\label{sec:results}

The performance measurements employ two AGC analysis setups to test different scalability issues: many-core scalability and  distributed scaling. Two different facilities were used as testbeds, one at the University of Nebraska–Lincoln, a second at the University of Chicago.

The University of Nebraska–Lincoln hosts a US-CMS Tier-2 site as well as resources dedicated to IRIS-HEP for development of the Coffea-Casa CMS analysis facility. This facility was used for an AGC scale-out performance benchmarking setup, with available resources including 12x Dell R750 each with dual Xeon Gold 6348 28C/56T CPUs, 512GB RAM, 200Gb networking, and 10x 3.2TB NVMe, providing in total 672 cores.

The facility at the University of Chicago was used to to test the scaling performance when using locally available input files. The university hosts a US-ATLAS Tier-3 site and additional resources dedicated to IRIS-HEP for the development of a Coffea-Casa ATLAS analysis facility. Resources available for testing included 16 nodes with dual Xeon Gold 6348 56C/112T CPUs, 384 GB RAM, and 10x 3.2 TB NVMe, providing in total 1792 cores.

\subsection{AGC setup with dataset stored on local disks}

The goal of these performance tests was to check the multi-core scalability for an AGC setup using \textit{coffea} as analysis framework and processing the AGC dataset stored on local disks to avoid network overhead. For efficient scaling over multiple local cores we used Python futures via the \textit{FuturesExecutor} in \textit{coffea}.

\begin{figure}[!ht]
\begin{minipage}[c]{0.48\textwidth}
\includegraphics[width=\textwidth]{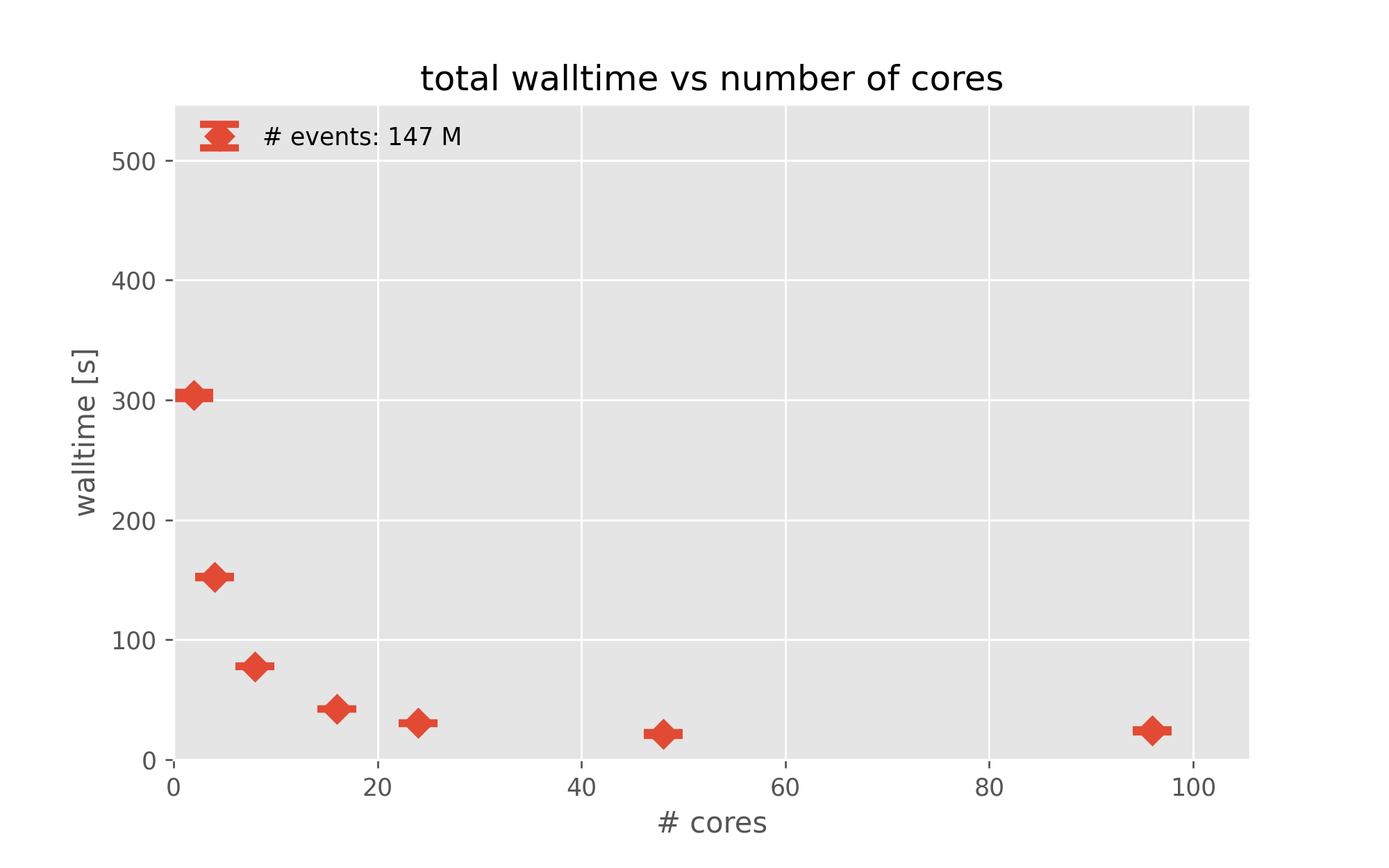}
\centering
\subcaption{Walltime measurements as a function of the number of cores used.}
\label{fig:perf1}
\end{minipage}\hspace{2pc}
\begin{minipage}[c]{0.48\textwidth}
\includegraphics[width=\textwidth]{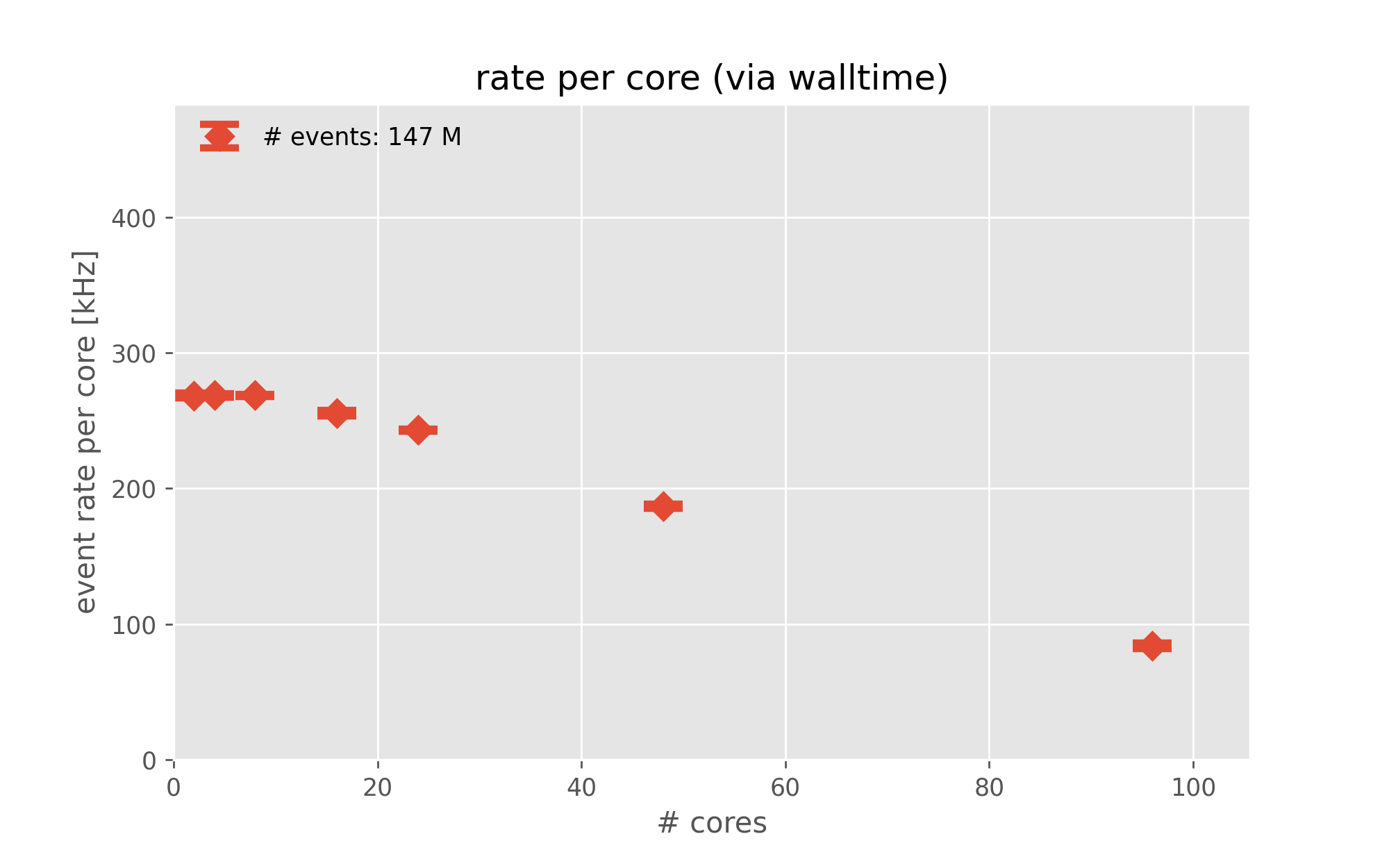}
\centering
\subcaption{Event rate as a function of the number of cores used.}
\label{fig:perf2}
\end{minipage} 
\caption{Reading locally stored files and scaling on local machine at the University of Chicago Coffea-casa AF.}
\label{fig:local-scaling}
\end{figure}

Figure~\ref{fig:local-scaling} shows the performance in terms of total walltime and event rate per core when scaling the analysis to use more cores. The slight degradation in efficiency may point towards some remaining overhead in the parallelization.

\subsection{AGC: scale-out to distributed resources}

Tests of the scale-out AGC implementation performance in a distributed setup were performed at the Coffea-casa facility at the University of Nebraska–Lincoln. We used a \textit{coffea} setup employing the \textit{DaskExecutor} to allow running tasks in a \textit{Dask}~\cite{dask} \textit{Distributed} cluster. The resulting jobs ran on the HTCondor Tier-2 batch queue.

\begin{figure}[!ht]
\begin{minipage}[c]{0.48\textwidth}
\includegraphics[width=\textwidth]{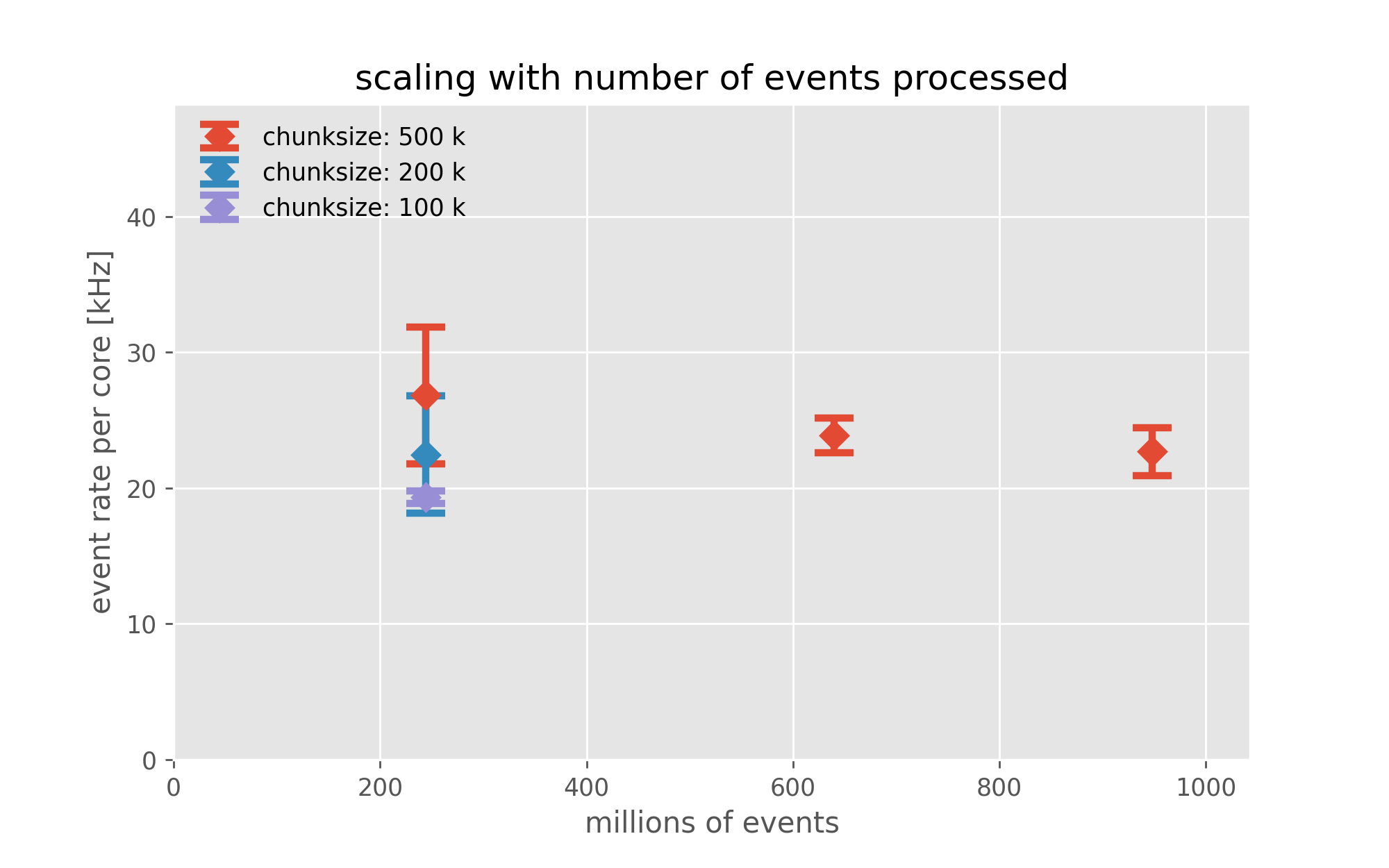}
\centering
\subcaption{Event rate scaling as a function of the number of events processed.}
\label{fig:perf3}
\end{minipage}\hspace{2pc}
\begin{minipage}[c]{0.48\textwidth}
\includegraphics[width=\textwidth]{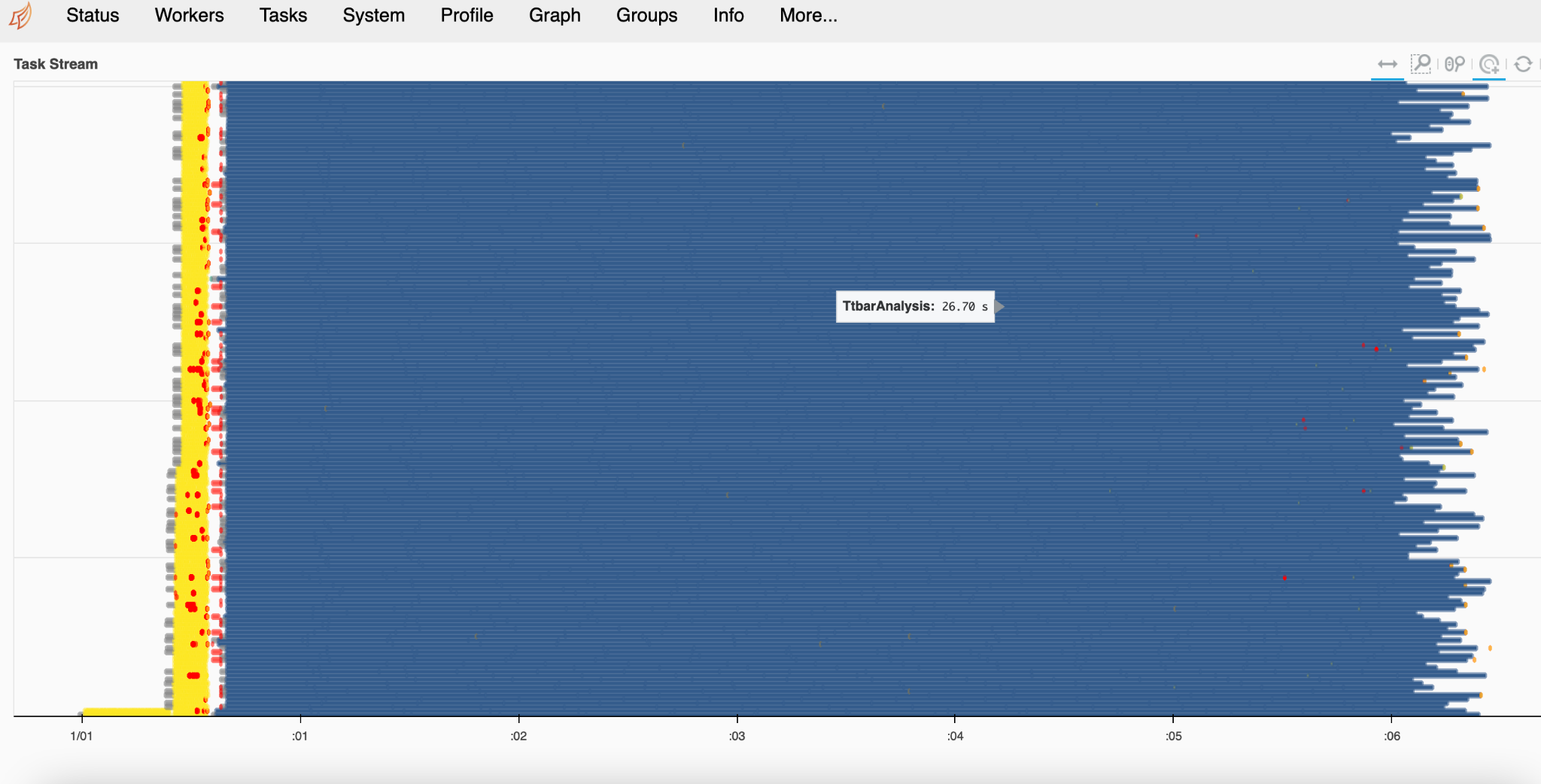}
\centering
\subcaption{Dask task graph showing efficient scheduling.}
\label{fig:taskgraph}
\end{minipage}
\caption{Using the Coffea-casa facility at the University of Nebraska–Lincoln CMS Tier-2 (\textit{coffea} with \textit{DaskExecutor}): stable scaling up to 1B events on the Tier-2 HTCondor job queue with efficient scheduling.}
\label{fig:distributed-scaling}
\end{figure}

Figure~\ref{fig:distributed-scaling} shows the event rates measured in scaling tests as a function of the number of events processed. This setup employs files read over the network. The event rate is stable, independent of the size of the dataset being processed. The \textit{Dask} task graph shows efficient scheduling of jobs performing the data processing.

\subsection{Scaling with I/O and number of cores}

Further scaling tests performed at the University of Nebraska–Lincoln Coffea-casa setup focused on the effect of using an increased number of cores and reading various fractions of the data in the input datasets.

\begin{figure}[!ht]
\begin{minipage}[c]{0.48\textwidth}
\includegraphics[width=\textwidth]{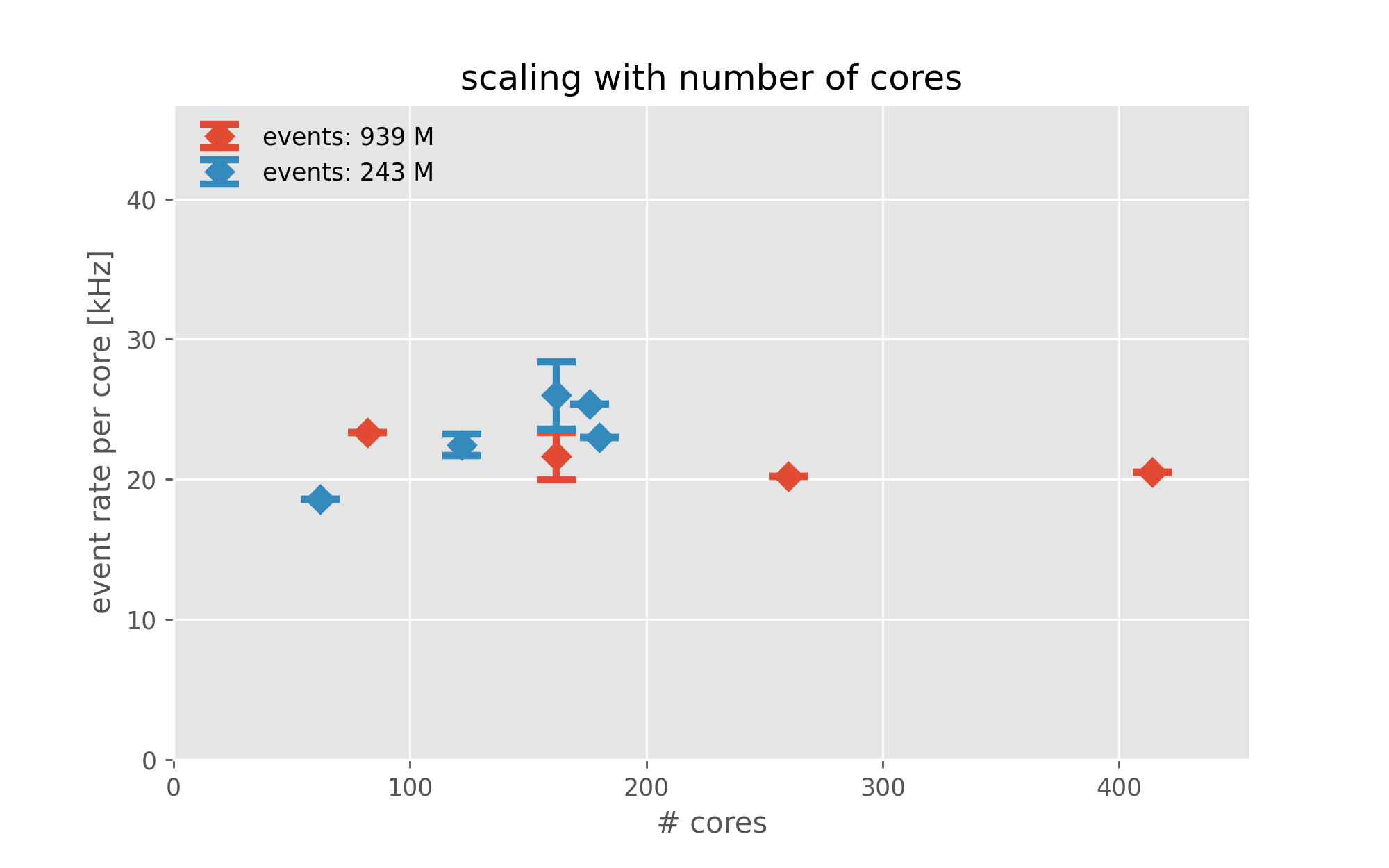}
\centering
\subcaption{Event rate scaling as a function of number of cores}
\label{fig:perf4}
\end{minipage}\hspace{2pc}
\begin{minipage}[c]{0.48\textwidth}
\includegraphics[width=\textwidth]{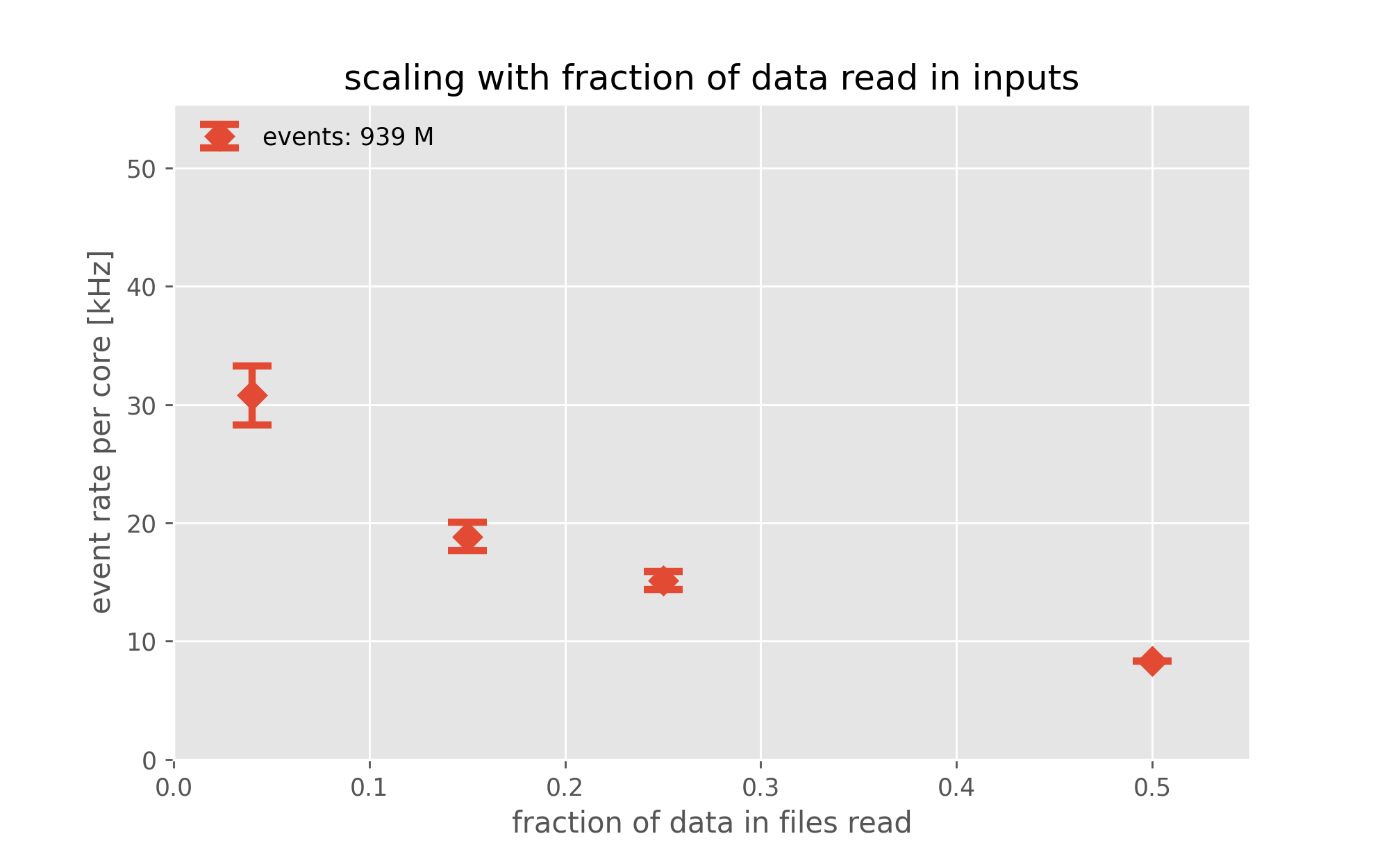}
\centering
\subcaption{Event rate scaling as a function of the fraction of data being read.}
\label{fig:perf5}
\end{minipage}
\caption{Using the Coffea-casa facility at the University of Nebraska–Lincoln CMS Tier-2 (\textit{coffea} with \textit{DaskExecutor}): stable scaling to 400 cores, event rates as a function of the fraction of data read.}
\label{fig:additional-scaling}
\end{figure}

Stable scaling is observed up to 400 cores as depicted in figure~\ref{fig:additional-scaling}. The fraction of data in the input files being read (which changes depending on the number of columns accessed in the file) can have a significant effect on the event rates, indicating that the time spent on data processing is not a significant contribution to the event rate when only reading a small fraction of the data in the files.

\section{Conclusion and outlook}
The first performance measurements obtained in the context of testing a specific pipeline implementing the AGC analysis task at various facilities show promising results. We plan to extend the analysis pipeline to include additional methods for data delivery and compare their performance in future work. We also expect to extend measurements to additional hardware configurations on various CMS and ATLAS analysis facilities.

\ack
This work was supported by the U.S. National Science Foundation (NSF) Cooperative Agreement OAC-1836650 (IRIS-HEP).
 
\section*{References}
\bibliography{references}

\end{document}